# Evolution of magnetic ordering in $FeCr_2Se_{4-x}Te_x$; x = 0 - 4.0

C. S. Yadav[1*], S. K. Pandey[2] and P. L. Paulose[3]

[1]School of Basic Sciences, Indian Institute of Technology Mandi, Mandi-175005, India
[2]School of Engineering, Indian Institute of Technology Mandi, Mandi-175005, India
[3]DCMP&MS, Tata Institute of Fundamental Research, Mumbai-400005, India

We have systematically studied the magnetic properties of chromium chalcogene compounds $FeCr_2Se_{4-x}Te_x$. The $FeCr_2Se_4$ undergoes antiferromagnetic ordering below 222 K. Substitution of tellurium lowers the antiferromagnetic ordering temperature and leads to short range ferromagnetic cluster behavior towards the tellurium end. Change over from antiferromagnetic to ferrimagnetic like behavior is also reflected in the corresponding transformation from semiconducting to metallic transport behavior. There is a large variation in the Curie-Weiss temperature, effective magnetic moment and ordering temperature ($T_N$ / $T_C$) with Te substitution. The electronic band structure calculations suggest antiferromagnetic and ferrimagnetic ground state for the $FeCr_2Se_4$ and $FeCr_2Te_4$ respectively.

## INTRODUCTION:

The ternary chromium chalcogenides $ACr_2X_4$ (A = Fe Co, Mn, Cu, Zn, Hg, Cd; X = S, Se) exhibit a variety of physical phenomena like metal-insulator transition, colossal magneto-resistance (CMR), and multiferroicity [1-3]. The $CuCr_2X_4$ (X = S, Se, Te) compounds are ferromagnetic metals with Curie temperature $T_C$ > 375 K, while $CdCr_2X_4$ and $HgCr_2X_4$ exhibit multiferroic properties with magneto-capacitance coupling [4-6]. $ZnCr_2S_4$ shows mixed ground state of spiral and collinear magnetic structures below 8 K owing to almost equal strength of ferromagnetic and antiferromagnetic (AFM) exchange [7, 8]. $ZnCr_2Se_4$ is ferromagnetic ($T_C$ ~ 100 K) dominated by spin fluctuations but finally undergo antiferromagnetic transition ($T_N$ ~ 21 K) [9].

The $FeCr_2X_4$ (X = S, Se, Te) compounds show competing spin-orbit and exchange interactions [3]. In $FeCr_2S_4$, ferrimagnetic order sets in below 165 K and shows CMR behavior [10-13]. The μ-SR studies have shown that the collinear ferromagnetic phase (T < 165 K) transforms into a non-collinear helical ordering below 60 K [10]. The magnetic transitions at 165 K and 60 K shifts towards higher temperature in the high pressure [11]. It also exhibits strong magnetocrystalline anisotropy and quadrupole splitting of Mossbauer spectra below Curie temperature [15]. The $FeCr_2Se_4$ is antiferromagnetic insulator with Neel temperature $T_N$ ~ 222 K [16-23]. It has monoclinic, brezinaite ($Cr_3S_4$) type structure where Fe atoms occupy ordered vacant sites in the metallic layer and Cr atoms resides in the layers without vacant sites [22]. The x-ray magnetic circular dichorism spectra of $FeCr_2Se_4$ shows strong hybridization between Fe 3d-states and Se p-states [16]. Such spectra points towards the presence of ferromagnetic component which is quite unusual for an antiferromagnetic (AFM) material. The x-ray absorption spectroscopy and photoemission spectroscopy measurements on $FeCr_2Se_4$ showed the valence states of Fe and Cr ions as divalent ($Fe^{2+}$) and trivalent ($Cr^{3+}$) respectively [16]. The structural and Mossbauer studies on the mixed-chalcogenide compounds $FeCr_2S_{4-x}Se_x$ show that the strength of the superexchange interaction between the neighboring moments remains unaffected by anion substitution in spite of the change in structure from cubic spinel ($FeCr_2S_4$) to monoclinic ($FeCr_2Se_4$) [23].

Though there are detailed electronic and magnetic studies on $FeCr_2X_4$ (X = S, Se) in literature, the $FeCr_2Te_4$ has not been studied much [24-26]. Andre et al. have reported the Fe-Cr-Te system near the $FeCr_2Te_4$ composition in the single crystals grown using vapor transport and the Bridgeman method [24]. We believe that the difficulty in preparing $FeCr_2Te_4$ in clean single phase might have been an obstacle in this regard. In the light of the interesting properties of $FeCr_2Se_{4-x}Te_x$, it is quite interesting to explore the effect of anion size effect on the physical properties of iso-structural $FeCr_2Se_{4-x}Te_x$. We have synthesized $FeCr_2Se_{4-x}Te_x$ series to study the evolution of structural and magnetic properties upon tellurium substitution. The crystal structure of $FeCr_2Se_{4-x}Te_x$ remains monoclinic (space group: $I_{2/m}$) in the full doping range. We observed that AFM transition gets suppressed upon Te substitution and a short range ferromagnetic behavior evolves for x ≥ 3.3. The end member $FeCr_2Te_4$ shows cluster glass like magnetic ordering below 140 K. The low temperature Mössbauer study reveals wide distribution of magnetic hyperfine field confirming the inhomogeneous magnetic ordering in $FeCr_2Te_4$.

## EXPERIMENTAL DETAILS:

The polycrystalline compounds of $FeCr_2Se_{4-x}Te_x$ (x = 0, 0.5, 1.0, 2.0, 3.0, 3.3, 3.6, 3.8, and 4.0) were prepared by the solid state chemical reaction of high purity constituent elements inside evacuated quartz tube in the temperature range 950 - 1020 °C. The x-ray diffraction measurement were done at room temperature using Philips x-ray diffractometer. The magnetic measurements were performed in the temperature range of 1.8 - 350 K, using Quantum Design SQUID magnetometer and Oxford Vibrating Sample Magnetometer. A Quantum Design PPMS was used for electrical resistivity measurements. $^{57}$Fe Mössbauer spectra was recorded in transmission geometry using a conventional constant-acceleration spectrometer and a helium flow cryostat. We have performed the electronic band structure calculation using WIEN2K program using Local Density Approximation (LDA) model [27].

**RESULTS AND DISCUSSION:**

The Rietveld refinement of the x-ray diffraction (XRD) data was performed using GSAS and FullProf programs [28-30]. The compounds near tellurium end required higher sintering temperature of 1020 °C instead of 950 °C in order to get rid of the impurity of FeTe. We have shown a representative XRD pattern of the FeCr$_2$Te$_4$ in the figure 1a. Crystal structure was refined in the monoclinic Cr$_3$S$_4$ like structure with space group: I$_{2/m}$. The lattice parameter, atomic position and parameter for goodness of fitting for FeCr$_2$Te$_4$ are shown in table 1. The variation of lattice parameters a, b, c, β and volume of the unit cell with tellurium concentration is shown the figure 1b. The error in the fitted value of lattice parameters is in fourth decimal place, therefore there is no error bar in the figure 1b. As seen from the figure the lattice parameters a, and b increase linearly from FeCr$_2$Se$_4$ to FeCr$_2$Te$_4$, whereas c parameter first increases for $0 \leq x \leq 3.0$ and then decreases for $3.3 \geq x \geq 4.0$. However the volume of the unit cell increases linearly due to the larger increase in a, and b compared to c. It is to note that the overall variation in c is ~ 1 % only. The angle β remains close to 90 $^0$ for x ≥ 1.0.

**Table 1:** Results of refinement parameters of x-ray diffraction for FeCr$_2$Te$_4$; [ I$_{2/m}$; Fe(2a), Cr(4i; (u, 0, v), S1(4i; (u, 0, v)), Se2 (4i; (u, 0, v)]

| Parameter | Value |
|---|---|
| a | 6.8143 Å |
| b | 3.9268 Å |
| c | 11.9429 Å |
| u(Cr) | -0.02913 |
| v(Cr) | 0.25944 |
| u(Se1) | 0.32869 |
| v(Se1) | 0.37420 |
| u(Se2) | 0.33993 |
| v(Se2) | 0.87658 |
| Rp (%) | 6.75 |
| wRp (%) | 9.96 |
| $\chi^2$ | 1.834 |

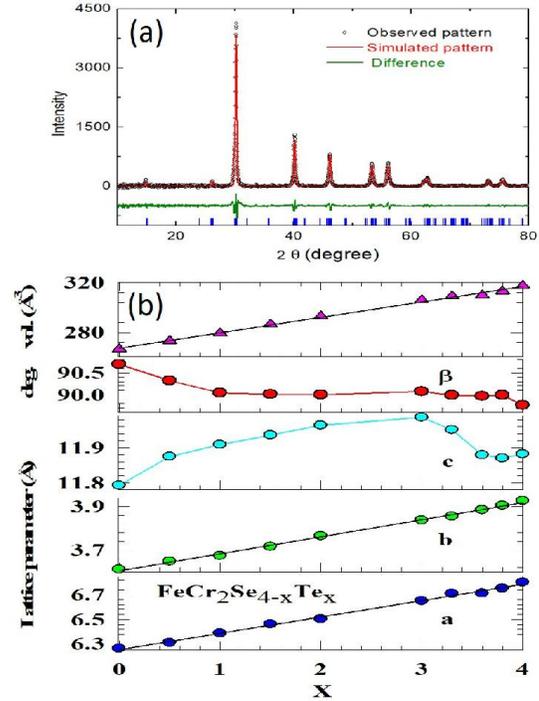

Figure 1. (Color online) (a) Rietveld refined x-ray diffraction pattern of FeCr$_2$Te$_4$, (b) The dependence of lattice parameters a, b, c, β and V on tellurium concentration.

The magnetization (M) versus temperature data of the FeCr$_2$Se$_{4-x}$Te$_x$ compounds measured at H = 100 Oe is shown in the figure 2. Compounds close to selenium end in the series show a cusp in M(T) curve at the AFM ordering temperature. Substitution of tellurium suppresses AFM ordering temperature T$_N$ from ~ 222 K for FeCr$_2$Se$_4$ to ~ 152 K for FeCr$_2$Se$_{0.7}$Te$_{3.3}$. It is to note that though there is clear cusp at AFM transition, magnetization shows upturn at low temperature for x = 0.5, 1.0, 2.0 and 3. Low temperature neutron diffraction studies on FeCr$_2$Se$_4$ suggested that the spins of Fe and Cr have AFM coupling along c-axis but ferromagnetic (FM) coupling along b-axis [16]. The FM interaction show its dominance over AFM interaction at low temperature and ferrimagnetic like magnetization is observed. As seen from the curve, a clear bifurcation of zero field cooled (ZFC) and field cooled (FC) magnetization curves is observed for For x > 2.0. This shows the enhancement in the FM exchange coupling on Te doping. It is plausible that the increase in the unit cell volume affects the AFM coupling along c -axis more than the ferromagnetic coupling along b axis. For x > 3.3, there is fast buildup of FM interactions and AFM order is completely suppressed. The FeCr$_2$Te$_4$ (i.e. x = 4.0) shows a clear peak at ~ 112 K with large divergence between ZFC and FC data. We measured M(T) at H = 0.5 T also for all the samples and carried out the Curie-Weiss fit in the temperature range above T$_N$/T$_C$ to 350 K. The Curie-Weiss parameters θ$_p$ and effective magnetic moment $\mu_{eff}$ obtained from the fit are shown in figure 3. The value of $\theta_p \sim -800$ K for $FeCr_2Se_4$ is comparable to the -751 K reported by Goya et al. [17] but much larger than -450 K by Kang et al. [20]. The $\theta_p$ value decreases rapidly and becomes positive towards tellurium end. The magnetic moment of $FeCr_2Se_4$, 8.3 $\mu_B$ is slightly higher than the 7.61 $\mu_B$ by Goya et al. [17]. The magnetic moment of the compounds decreases from 8.3 $\mu_B$ for $FeCr_2Se_4$ to 5.9 $\mu_B$ for $FeCr_2Te_4$. The isothermal magnetization M(H) measured at 2 K (fig. 4) shows hysteresis for the compositions corresponding to x = 4.0, 3.8, and 3.6; without the tendency of saturation of magnetic moments even at 12 Tesla field. For x $\leq$ 3.3, M(H) curve is almost linear as expected for AFM systems. This behavior further shows dominance of Ferrimagnetic exchange interaction for x $\geq$ 3.8 and AFM ordering for x $\leq$ 3.3.

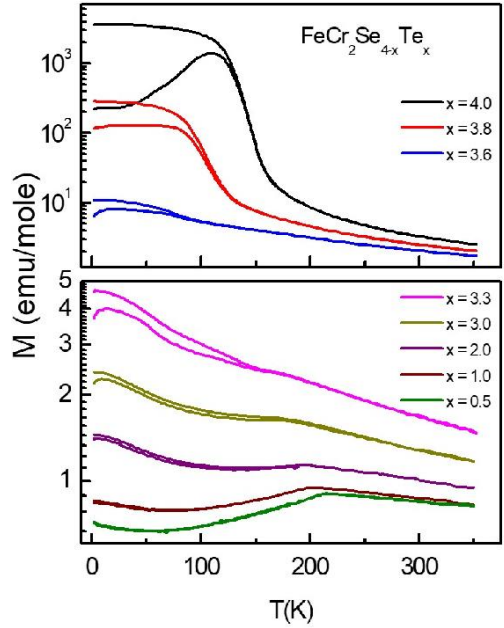

Figure 2 (Color online). M versus T for $FeCr_2Se_{4-x}Te_x$ (x = 0.5, 1.0, 2.0, 3.0, 3.3, 3.6, 3.8, and 4.0), at H = 100 Oe.

We have shown the variation of magnetic transition with tellurium doping in figure 3. The AFM coupling that sets in at 222 K for $FeCr_2Se_4$ is quite robust up to ~ 80 % replacement of selenium by tellurium (x = 3.3) and $T_N$ decreases to ~ 152 K for $FeCr_2Se_{0.7}Te_{3.3}$. For higher tellurium concentration (x $\geq$ 3.3), ferrimagnetic cluster state emerges at lower temperature ($T_G$ ~ 30 K) for $FeCr_2Se_{0.4}Te_{3.6}$. The state strengthens further for higher tellurium concentration (T~ 112 K) for $FeCr_2Te_4$. It is notable that there is reduction in the length of unit cell along c-axis for x > 3.0. The reduction in c is likely to enhance hybridization between d-orbitals of Fe and Cr atoms with p-orbitals of chalcogene and hence may play significant role in magnetic coupling between the transition metal atoms.

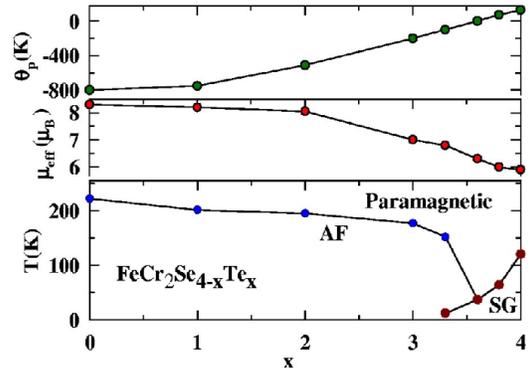

Figure 3 (Color online). The Curie Weiss temperature $\theta_P$, magnetic moment per formula unit ($\mu_{eff}$) and the ordering temperature ($T_N/T_C$) are plotted as a function of the tellurium concentration.

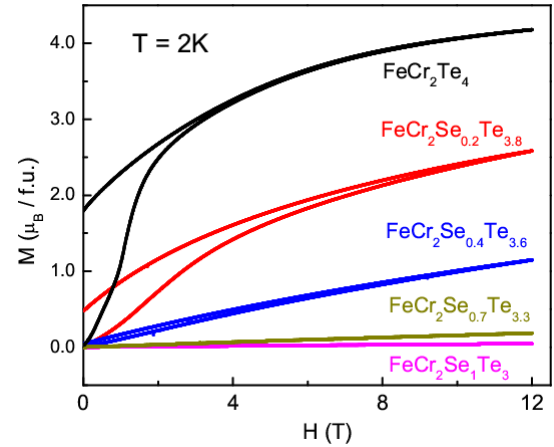

Figure 4 (Color online): M versus H curves taken at T = 2 K for $FeCr_2Se_{4-x}Te_x$ (x = 3.3, 3.8, 3.6, 4.0).

In this section we discuss the results of magnetic measurements on $FeCr_2Te_4$. The M(T) curves taken at H = 100 Oe field are shown in the figure 5(a). Magnetization shows sharp increase below 140 K like the ferromagnet or ferrimagnetic compound. However, the bifurcation of ZFC and FC curves below 112 K, indicates the formation of magnetic cluster units. The Curie-Weiss fit of data with $\chi(T) = \frac{C}{T-\theta_p}$ in temperature range 160 - 350 K yields $\theta_p \sim$ 140 K and $\mu_{eff}$ of ~ 5.9 $\mu_B$ per formula unit. The isothermal magnetization measured at 2 K (figure 5(b)) does not fully saturate up to 12 Tesla and shows a weak increase on increasing magnetic field. It is to note that the M(H) shows magnetic hysteresis arising from the dominating ferrimagnetic state.

The AC susceptibility measurements (figure 5(c)) performed at frequencies f = 0.12, 1.2, 12, 120, and 1200 Hz show shift in peak position towards higher temperature on increase in frequency. For the frequency variation of over 4 decades, freezing temperature $T_f$ is reduced by about 1 %. A quantitative measure of the frequency shift is obtained by the expression [31-33]; $\delta T_f = \frac{\Delta T_f}{T_f \delta(\log \nu)}$. It is interesting to note that the obtained value of $\delta T_f$ is ~ 0.004 which is comparable to ($\delta T_f = 0.005$) for the canonical spin glass system CuMn [33]. Additionally, we tried to fit the susceptibility data with the Vogel-Fulcher relation [29]; $\ln\frac{\nu}{\nu_0} = -\frac{E_a}{k_B(T_f - T_0)}$, where $E_a$ is the potential barrier separating two adjacent magnetic clusters, $\nu_0$ is the characteristic frequency of the clusters, $T_f$ is the freezing temperature and $T_0$ is temperature, which is a measure of the inter-cluster interaction strength. The inset of the Fig 5(c) shows the fit to the $T_f$ versus log ($\nu$) data, which gives $E_a/k_B$ = 78.7 K, $\nu_0 = 5.2 \times 10^{10}$ Hz, and $T_0 = 108.9$ K. The non-zero value of $T_0$ suggests finite interaction between spins and formation of the magnetic clusters. We have measured M versus T at hydrostatic pressures of P = 0, 3, 5, 7, 9 kbar on FeCr$_2$Te$_4$. As seen from the figure 5(d), ordering temperature shifts towards lower temperatures upon application of pressure. This result coincides with the fact that ordering temperature also shifts towards lower temperature for FeCr$_2$Se$_{0.2}$Te$_{3.8}$ and FeCr$_2$Se$_{0.3}$Te$_{3.7}$ that have smaller volume in comparison to FeCr$_2$Te$_4$.

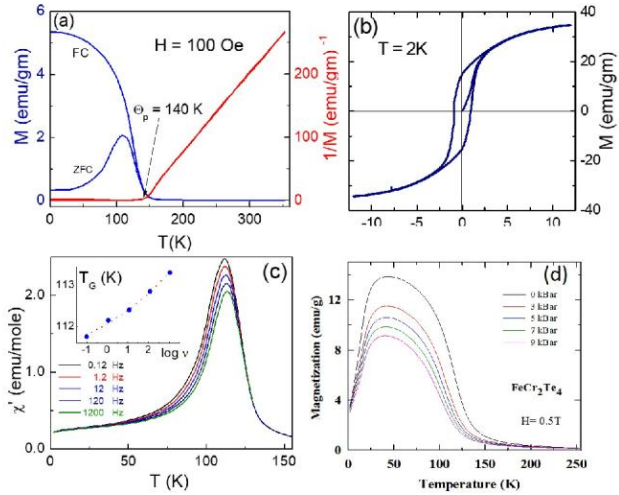

Figure 5 (Color online): (a) DC magnetization of the FeCr$_2$Te$_4$ plotted as M versus T and 1/M versus T measured at H = 100 Oe. (b) Magnetic hysteresis curve for FeCr$_2$Te$_4$ taken at 2 K shows hysteresis behavior but does not saturate up to H = 12 T. (c) Temperature and pressure dependence of dc magnetization measured at H = 5 kOe. (d) AC susceptibility χ of FeCr$_2$Te$_4$ measured at different frequencies, shows spin glass like behavior below 140 K. The inset shows the linear fit to the Vogel-Fulcher law.

We have carried out $^{57}$Fe Mössbauer spectroscopy studies to probe that local magnetic state of Fe in FeCr$_2$Te$_4$. The experimental spectra were least-square fitted to get the hyperfine parameters namely isomer shift (IS), magnetic hyperfine field ($H_{hf}$) and quadrupole moment. The Mössbauer spectra at 295 K, (Fig.6) shows a quadrupole doublet ($e^2qQ/2 = 0.33$ mm/s) with a line width 0.39 mm/s, typical of paramagnetic state. The isomer shift is 0.57 mm/s and it shows that Fe is in 2$^+$ state. Noticeable broadening of the spectral line is observed at 101 K in agreement with the bulk magnetization data. The Hyperfine field continues to evolve as the temperature is decreased. We could fit the 5 K spectra using seven independent sub-spectra with ($H_{hf}$) in the range of 2.5 Tesla to 21 Tesla and $e^2qQ/2 < 0.33$ mm/s. The spectra at 101 K and 51 K was fitted using 3 and 5 sub-spectra respectively. The fit reveals that about 33% of the Fe nuclei experience negligible hyperfine field at 101 K whereas at 51 K, all the Fe nuclei attain a ($H_{hf}$) in the range 2.6 Tesla to 13 Tesla. The probability distribution of hyperfine field at 5 K is shown in Figure 6(b). The distribution indicates the sensitive nature of the FE hyperfine field to its local environment and an inhomogeneous magnetic state. The other notable feature is that the maximum hyperfine field is only 21 Tesla (85 % of $H_{hf}$ is less than 12.6 Tesla) indicating a substantially low magnetic moment on Fe ion.

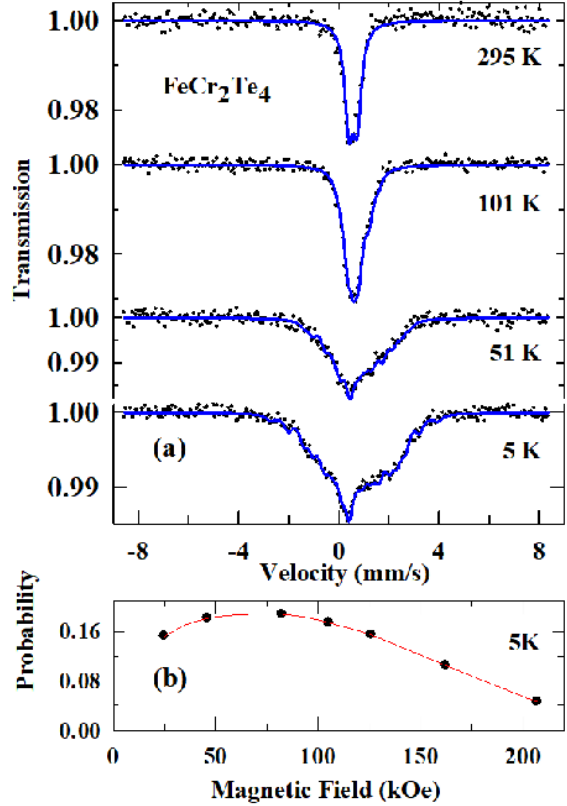

Figure 6 (Color online). Mössbauer spectra of FeCr$_2$Te$_4$ at 295 K, 101 K, 51 K and 5 K. (b) Hyperfine field distribution from the Mossbauer data at 5K.

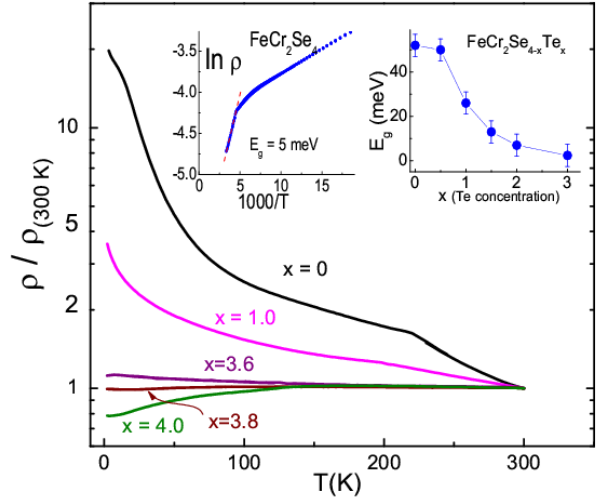

Figure 7 (Color online). Normalized electrical resistivity of $FeCr_2Se_{4-x}Te_x$; x = 0, 1.0, 2.0, 3.6, 3.8, 4.0. The inset shows (a) Arrhenius plot for $FeCr_2Se_4$ (b) Variation of $E_g$ with doping concentration.

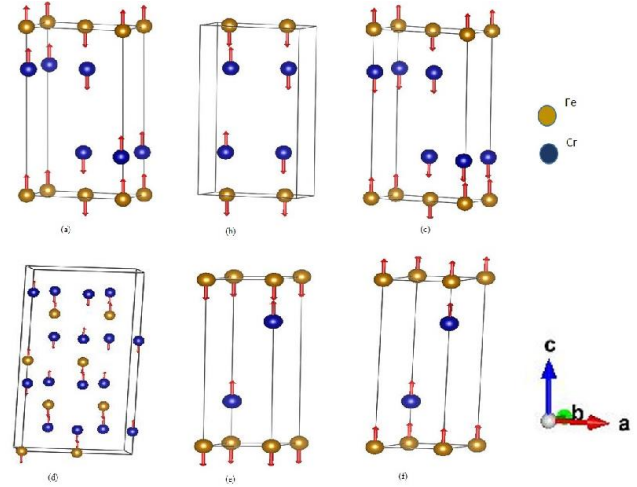

Figure 8 (Color online). Various possible spin configurations; (a) AFM-1, (b) AFM-2, (c) AFM-3, (d) AFM-4, (e) Ferri, and (f) Ferro; of Fe and Cr atoms for the $FeCr_2Se_4$ and $FeCr_2Te_4$ compounds. For AFM-4 configuration, a 2×1×2 cell is shown, other configuration are shown for primitive cell.

Based on the above mentioned studies, it is evident that the crystal structure of the $FeCr_2Se_{4-x}Te_x$ remains monoclinic throughout the doping range. Though a, b, and volume of the unit cell monotonously increases, the c-axis lattice parameter shows decrease for x > 3.0. The increase in a and b upon tellurium doping, results in the reduced interaction in Fe-Fe atoms and Cr-Cr atoms along their respective planes. This might be responsible for the gradual decrease in AFM ordering temperature on increase of volume of the unit cell. However the reduction in c-axis for x > 3.0 indicates the possibility of enhanced hybridization between d-orbitals of Fe and Cr atoms with the p-orbital of selenium/tellurium. The magnetic ordering also follows similar trend as $T_N$ drops sharply for x > 3.3 and short range ferromagnetic order appears simultaneously. The onset of this short range ferromagnetic order for x > 3.3 compositions coincide with the reduction in c -axis value. The effective magnetic moment is also altered towards tellurium rich end. The Mossbauer spectroscopy shows a variation of Fe hyperfine fields from 2.5 Tesla to 21 Tesla indicating inhomogeneous Fe moment distribution. All these results are in agreement with the band structure calculations predicting a ferrimagnetic ground state and a reduced Fe moment for $FeCr_2Te_4$ as we move from $FeCr_2Se_4$.

We have carried out the electrical transport studies across the whole range x = 0 - 4.0. The compounds show metallic behavior (Figure 7) for x > 3.6, which coincides with the occurrence of ferromagnetic clustering as observed in magnetization data. For x ≤ 3.0, compounds show semiconducting behavior. We have shown ln ρ versus 1/T in the inset (a) to calculate the activation energy. The obtained value of activation energy ($E_g$) for all the compounds in the high temperature range (shown in the inset b) decrease sharply upon Te substitution. To understand the magnetic behavior of compounds we performed LDA based electronic structure calculations and found the magnetic and non-magnetic solutions for $FeCr_2Se_4$ and $FeCr_2Te_4$. We observed that the magnetic state of both compounds has lower energy than the non-magnetic state. The obtained energy differences between magnetic and non-magnetic ground states of $FeCr_2Se_4$ and $FeCr_2Te_4$ are 1.313 eV and 0.755 eV respectively. This energy difference can be used to roughly compare the magnetic ordering temperatures in the compounds that indicates the ordering temperature for $FeCr_2Se_4$ to be ~ 1.7 times higher compared to that for $FeCr_2Te_4$. Such estimation is very close to the ratio ~ 1.6, obtained from the experimental values of the magnetic ordering temperature of 222 K and 140 K for $FeCr_2Se_4$ and $FeCr_2Te_4$ respectively.

Further, considering the Fe and Cr spins parallel to each other in primitive unit cell, the calculated magnetic moments per formula unit for $FeCr_2Se_4$ and $FeCr_2Te_4$ are 8.44 $\mu_B$ and 8.26 $\mu_B$ respectively. The respective contributions of Fe, Cr and interstitials magnetic moments to the total moment were obtained as 2.64 $\mu_B$, 2.64 $\mu_B$ and 0.63 $\mu_B$ for $FeCr_2Se_4$; and 2.33 $\mu_B$, 2.74 $\mu_B$ and 0.64 $\mu_B$ for $FeCr_2Te_4$. The contribution of moment from interstitials is more for $FeCr_2Te_4$ than for $FeCr_2Se_4$. Since we have kept the muffin tin radii for Fe and Cr, same in both compounds, relative change in Fe and Cr moments for these compounds indicates the Fe-Te interaction to be more effective than Cr-Te interaction.

Table -2 Ground state energies (in meV) for different magnetic states.

| Magnetic state | $FeCr_2Se_4$ | $FeCr_2Te_4$ |
|---|---|---|
| AFM-1 | $\Delta_0 + 191$ | $\Delta_0 + 231$ |
| AFM-2 | $\Delta_0 + 179$ | $\Delta_0 + 206$ |
| AFM-3 | $\Delta_0 + 147$ | $\Delta_0 + 231$ |
| AFM-4 | $\Delta_0$ | $\Delta_0 + 55$ |
| Ferri | $\Delta_0 + 55$ | $\Delta_0$ |
| Ferro | $\Delta_0 + 278$ | $\Delta_0 + 84$ |

If we make an approximation of random alignment of magnetic moments of the primitive unit cells (keeping the spin alignment within the primitive unit cell intact); calculated moment value of 8.44 $\mu_B$ for $FeCr_2Se_4$ corresponding to s = 4.2 $\mu_B$, translates into effective magnetic moment $(2\sqrt{s(s+1)})$ value ~ 9.35 $\mu_B$, which is comparable to the experimentally measured $\mu_{eff}$ ~ 8.3 $\mu_B$. However similar estimation for $FeCr_2Te_4$ gives theoretical $\mu_{eff}$ values as 9.21 $\mu_B$, which is larger than the experimental value 5.9 $\mu_B$. Furthermore, such approximation gives AFM ground state for both compounds which is not true in case of $FeCr_2Te_4$. Therefore we considered various spin configurations (shown in Figure 8) within the primitive unit cell and calculated the ground state energy of these configurations. The ground state energies of the compounds for these spin configurations are shown in table 2. The $FeCr_2Se_4$ was found to have minimum ground state energy for AFM-4 configuration, where Fe and Cr spins are parallel to each other in the primitive unit cell, and these cells are antiferromagnetically coupled to each other in the structure. Bin et al. have also found the low energy antiferromagnetic ground state for this type of magnetic cell structure [16]. On the other hand, for $FeCr_2Te_4$ we obtained ferrimagnetic configuration (Fe and Cr spins antiparallel to each other) to have minimum ground state energy for the primitive cell with ferromagnetic coupling between these primitive cells. The calculated magnetic moment of ~ 2.7 $\mu_B$ for $FeCr_2Te_4$ ground state corresponds to the saturation magnetization moment of 1.35 $\mu_B$, which is much smaller than experimentally observed 3.2 $\mu_B$ value taken at the point of the closure of hysteresis at H = 4 T field. Although the LDA calculation is able to generate true ground state for $FeCr_2Te_4$, it underestimates the value of the magnetic moment.

**CONCLUSIONS:**

We have systematically studied the magnetic properties of $FeCr_2Se_{4-x}Te_x$. The substitution of Te lowers antiferromagnetic ordering temperature and leads to the onset of short range ferromagnetic ordering towards the tellurium end. The sharp reduction in the long axis (c -axis) lattice parameter close to tellurium end (x ≥ 3.6) and high pressure magnetic studies corroborate well with the development of ferromagnetic exchange in the compound. Our results shows a change in the Curie Weiss temperature $\theta_p$ and magnetic moment $\mu_{eff}$ for x ≥ 2.0 compounds. Hyperfine field distribution obtained from Mössbauer spectroscopy also proves the existence of inhomogeneous magnetic order at low temperature for $FeCr_2Te_4$. The LDA based electronic band structure calculations also indicate ferrimagnetic order.

**ACKNOWLEDGEMENT:** We thank Manish Ghag for the help during the different stages of sample preparation and measurement. CSY acknowledges AMRC IIT Mandi for the experimental facility and institute seed grant project IITM/SG/ASCY/29 for the financial assistance.